\documentclass[twocolumn,prb,showpacs]{revtex4}
\usepackage{psfrag}
\usepackage{graphicx}
\usepackage{amsmath}
\usepackage{amssymb}

\begin{document}

\title{Conformation dependence of charge transfer and level alignment in nitrobenzene junctions with pyridyl anchor groups}
\author{R. Stadler}
\affiliation{Department of Physical Chemistry, University of Vienna, Sensengasse 8/7, A-1090 Vienna, Austria}

\date{\today}

\begin{abstract}
The alignment of molecular levels with the Fermi energy in single molecule junctions is a crucial factor in determining their conductance or the observability of quantum interference effects. In the present study which is based on density functional theory calculations, we explore the zero-bias charge transfer and level alignment for nitro-bipyridyl-phenyl adsorbed between two gold surfaces which we find to vary significantly with the molecular conformation. The net charge transfer is the result of two opposing effects, namely Pauli repulsion at the interface between the molecule and the leads, and the electron accepting nature of the NO$_2$ group, where only the latter which we analyze in terms of the electronegativity of the isolated molecules depends on the two intra-molecular torsion angles. We provide evidence that the conformation dependence of the alignment of molecular levels and peaks in the transmission function can indeed be understood in terms of charge transfer for this system, and that other properties such as molecular dipoles do not play a significant role. Our study is relevant for device design in molecular electronics where nitrobenzene appears as a component in proposals for rectification, quantum interference or chemical gating.
\end{abstract}
\pacs{73.63.Rt, 73.20.Hb, 73.40.Gk}
\maketitle

\section{Introduction}

In the emerging field of single molecule electronics~\cite{molelect} considerable progress has been made in recent years in achieving comparable results for the electron transport through nano junctions from theory and experiment for a small range of well-defined test systems such as H$_2$ molecules between Pt leads~\cite{h2pt1}-~\cite{h2pt4} or bipyridine~\cite{bipy1}-~\cite{bipy5} and alkane mono- or di-thiolates~\cite{alkdithio1}-~\cite{alkdithio6} between Au electrodes. None of these systems, however, show interesting device properties and although a variety of theoretically designed single molecule devices~\cite{device1}-~\cite{device4}  exist in the literature, these proposals usually assume an idealized atomic configuration for the junction setup, where most degrees of freedom for the position of the nuclei are frozen and which can rarely be matched by experiments. A molecular component, which is particularly popular for molecular rectifiers~\cite{rect1}-~\cite{rect5} but also more recently has been suggested in the context of quantum interference effects~\cite{qie1}-~\cite{qie4} is a conjugated or cross-conjugated $\pi$ system substitued with one or more NO$_2$ groups. 

It is by now well established that the alignment of the eigenenergies of molecular orbitals (MOs) with the Fermi energy of the metal electrodes~\cite{scandolo}-~\cite{heimel2}, which is a key factor in determining the conductance of a molecular nano junction, is strongly related to the zero-bias charge transfer between the molecule and the leads~\cite{fermimy1}-~\cite{heimel3} where also the effect of Pauli repulsion at the interface (often addressed as interface dipole or pillow effect) can play a role~\cite{fermimy1,fermimy2}. Hence, the electron accepting nature of nitro-substituents is routinely used to induce an upwards shift of molecular levels~\cite{qie2} which is also supported by theoretical trend studies for ideal configurational setups of the junction geometry~\cite{shift1,shift2}. Such shifts have even been used to propose the concept of chemical gating~\cite{hybertsen1}, where data has been presented for a variety of electron accepting and donating groups but NO$_2$ has not been included in the study. In a recent work based on photoemission spectroscopy~\cite{ratner}, however, empirical evidence was found, that the assumed strong correlation between the donor or acceptor strength of chemical substituents on the one side, and level alignment or work function reduction (in the case of adsorbed monolayers) on the other side does not always hold, and hints at a dependence of this correlation on molecular orientation were given~\cite{ratner}. While the dependence of the electronic coupling on molecular conformation in biphenyl with rather strong coupling to gold leads has been investigated recently both by experiment~\cite{hybertsen,mayor} and theory~\cite{rect5,pauly,lambert}, no attention has been paid so far to the effect of finite torsion on level alignment, which is more decisive for the conductance in systems with weaker bonds to the electrodes as it was found in a comparison between bipyridine and biphenyl-dithiolate~\cite{fermimy1,fermimy2}.

In the present work we show in calculations based on density functional theory (DFT) and employing the Siesta code~\cite{siesta} that for nitro benzene anchored to Au electrodes by pyridyl linkers (Fig.~\ref{fig1}a) (a system we recently investigated in the context of quantum interference effects~\cite{qie4}) the molecule loses fractions of electrons to the leads for a wide range of geometries in spite of the electron accepting nature of the NO$_2$ group. This finding is explained by i) relatively low values for the electron affinity (EA) of nitro-bpph and ii) the influence of Pauli repulsion effects at its interface to the leads as a mechanism for electron removal from the molecule which we explored in great detail in earlier work for the Au/bipyridine/Au system~\cite{fermimy1,fermimy2}. Although the values for EA are rather low, their conformation dependence still determines the dependence of charge transfer on the two torsion angles $\phi_1$ and $\phi_2$ in Fig.~\ref{fig1}a, where we have studied 100 different atomic configurations for the molecule in the junction, while Pauli repulsion at the interface is found to be conformation independent. We also present evidence suggesting that in this junction setup the dependence of the molecular level alignment on $\phi_1$ and $\phi_2$ is dominated by the variation of charge transfer (which includes interfacial Pauli repulsion in our definition) and that for this dependence the molecular dipole and changes in the composition of MOs should play a rather insignificant role.

The paper is organized as follows: In the next section we describe our computational methods and introduce the scope of molecular conformations studied, where we also address the question of their respective stability. The following section contains a discussion of our results in five steps. First we establish the conformation dependence of transmission functions and molecular eigenenergies in Au/nitro-bpph/Au junctions which is the main motivation of our study and also its link to our previous work in Ref.~\cite{qie4}. Secondly, we discuss equilibrium charge transfer and its driving forces for the single molecule junctions in our article, where the effects of both interfacial Pauli repulsion and molecular electronegativity are discussed. In a third step we argue that charge transfer explains the conformation dependence of the alignment of molecular levels rather than molecular dipoles or a change of composition in MOs. We then check explicitly the impact of shortcomings of our DFT approach on the conclusions in this article by applying {\it ad hoc} corrections for the underestimation of energy gaps and the methods failure to describe dynamical screening effects. Finally, we describe the changes in molecular dipole moments with the two torsion angles $\phi_1$ and $\phi_2$. The last section contains a summary of our main conclusions. 

\section{Methods and structures}

\subsection{Computational details}\label{sec:comp}

Throughout this article we have generated our data by using the density functional theory (DFT) based Siesta code~\cite{siesta} with a double-zeta polarized (DZP) basis set, Troullier-Martins type norm-conserving pseudopotentials~\cite{martins} and a Perdew-Burke-Ernzerhof (PBE) parametrization~\cite{perdew2} for the exchange correlation (XC) functional. For the electron transport problem a non-equilibrium Green's function (NEGF) approach~\cite{kristian} has been employed which in the details of its combination with Siesta has been described in Ref.~\cite{mikkel}. In order to apply this method it is necessary to divide single molecule junctions into three regions, namely a left lead, a right lead and a scattering region. For all three regions independent supercell calculations with periodic boundary conditions are performed, where in the plane perpendicular to the transport direction we chose 3x3 sections of Au fcc (111) to form the unit cell for all geometries in this article. In a second step the leads are added to the Green's function of the scattering region as self energies allowing for the calculation of transmission functions within the framework of the Keldysh formalism~\cite{keldysh}. For the {\bf k}-point sampling in the transverse plane a 4x4 grid has been used for all systems but it has been checked for one structure that 6x6 and 8x8 grids do not change our results in any way.

A possible source of concern might be seen in our LCAO basis set which was shown recently to give inaccurate results for various properties of noble metal surfaces such as surface energies and work functions~\cite{ordejon}. The reason for this deficiency was identified in the too rapid decay of the surface wavefunctions into the vacuum. This problem results only in slight errors of $\sim$ 0.1 eV for surface energies but nevertheless can be dramatic in terms of percentages because the latter quantity is generally small for noble metals. The work function is found to be underestimated by $\sim$ 1 eV by a DZP basis set for gold in Ref.~\cite{ordejon}, since it is calculated directly from the electrostatic potential in the vacuum which naturally depends very strongly on the quality of the description of the long-range decay of the wavefunction. Most importantly for our study, the DZP-LCAO wavefunctions were only seen to deviate substantially from accurate plane wave calculations beyond a distance of 2-3 \AA \ . In our work half the bonding distance between the molecule and the leads must be seen as the relevant length scale for wavefunction decay regarding e.g. the accuracy of interface dipoles. This length is 1.06 \AA \ and we therefore expect our main results to be unimpaired by the named difficulties. We concede, however, that these technical problems affect our description of the free gold surface and the decay of its electron density into the vacuum, which can have an influence on the absolute values for the transfer of fractional charges which we discuss in Sec.~\ref{sec:charge}. But such an error is clearly independent of molecular conformation since the free Au surface is always the same for all systems in this article and therefore has no relevance for our discussion. We also stress that we use trends in fractional charges only as a means of elucidating the physical driving forces for level alignment and that such fractions of charges are not themselves observable quantities which could be measured in experiments directly.

\subsection{Molecular junctions and their stability}\label{sec:stab}

\begin{figure}
\includegraphics[width=1.0\linewidth,angle=0]{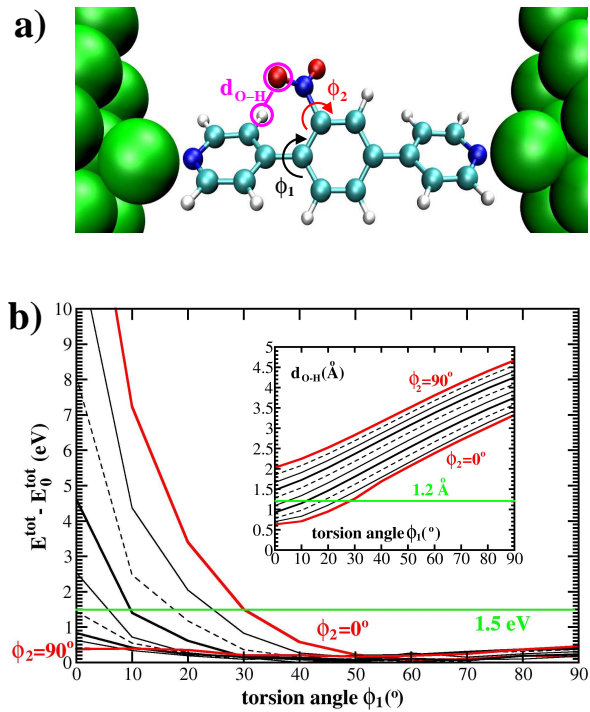}
      \caption[cap.Pt6]{\label{fig1}(Color online) a) Geometry of the Au/nitro-bpph/Au junctions in this article. b) Total energies for all angles $\phi_1$ and $\phi_2$ with respect to the most stable conformation ($\phi_1$=40$^{\circ}$, $\phi_2$=50$^{\circ}$). The torsion dependence of the distance d$_{O-H}$ between the H and O atoms causing the largest contribution to steric repulsion (which are marked in a) ) is also displayed as inset. The green lines give an indication which structures are stable or could be stabilized by chemical means, namely the ones with E$^{tot}$ less than 1.5 eV or d$_{O-H}$ above 1.2 \AA \ .  }
    \end{figure}

\begin{figure*}
\includegraphics[width=0.5\linewidth,angle=270]{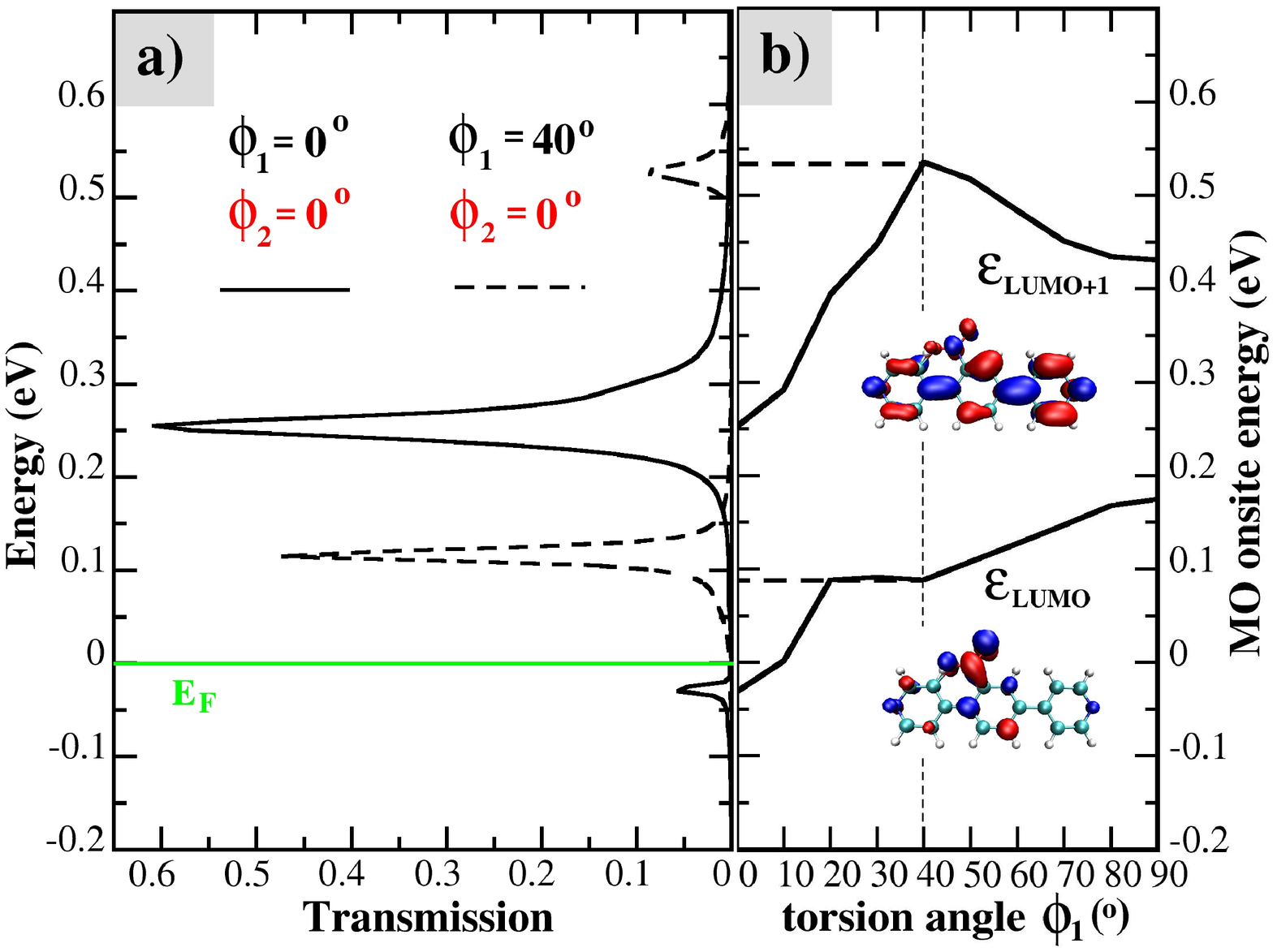}
\vspace{0.5 cm}
\caption[cap.Pt6]{\label{fig2}(Color online) a) Transmission functions at energies close to E$_F$ for two geometries with ($\phi_1$=0$^{\circ}$/$\phi_2$=0$^{\circ}$, solid line) and ($\phi_1$=40$^{\circ}$/$\phi_2$=0$^{\circ}$, dashed line). b) Evolution of MO eigenenergies with $\phi_1$ for the LUMO and LUMO+1 for $\phi_2$=0$^{\circ}$. The spatial distributions for these two orbitals are shown as insets for ($\phi_1$=0$^{\circ}$/$\phi_2$=0$^{\circ}$). }
\end{figure*}

Fig.~\ref{fig1}a shows the geometry of 2,5-bis(4-pyridyl)nitrobenzene coupled to Au (111) surfaces with an ad-atom where the N-Au distance on both sides is 2.12 \AA \ which has been identified as equilibrium distance for this system in earlier work~\cite{bipy3,qie4}. In Refs.~\cite{bipy3,qie4} the choice of this particular surface structure with an ad-atom has been motivated as due to its effect of bringing the lowest unoccupied MO (LUMO) close to the Fermi energy in the junction where a detailed comparison with flat surfaces has also been undertaken. Throughout this article the abbreviation nitro-bipyridyl-phenyl (nitro-bpph) has been used for the molecule for the sake of simplicity. The two torsion angles $\phi_1$ and $\phi_2$ have been varied in our calculations resulting in a total of 100 different molecular confomations. The atomic configuration of the molecule has been optimized once by DFT total energy minimization and then the angles have been varied rigidly. For the Au surfaces the atomic positions have been based on truncated bulk values. No relaxations of nuclei positions resulting from the interaction between molecule and leads have been considered in this article, because as can be argued from the rather small bonding energies for pyridil anchors on gold surfaces~\cite{bipy3}, such relaxations could only result in minor corrections but this would come at the expense of increasing the computational effort substantially.

A main purpose of our article is to investigate the influence of the torsion angles $\phi_1$ and $\phi_2$ in nitro-bpph (see Fig.~\ref{fig1}a) on charge transfer and level alignment. In this section we ask the question how this two degrees of freedom affect the stability of the molecule in the junction. For unsubstituted molecules consisting of directly connected aromatic rings such as biphenyl~\cite{biphenyl} and bipyridine~\cite{bipy3} it is well known that their tilt angles, $\sim$ 45$^{\circ}$ and $\sim$ 25$^{\circ}$, respectively, are a result of the interplay between the $\pi$ conjugation, which tries to flatten the structure, and the steric repulsion of the hydrogen atoms in the ortho position with respect to the ring-connecting carbons, which favors a perpendicular arrangement of the planes of two adjacent rings. By chemical means this balance has been shown to be adjustable to lower angles by introducing a -CH$_2$- bridge and to higher angles with bulky side-groups~\cite{hybertsen,mayor}. 

For nitro-bpph the situation is more complicated, partly because there are now two conformational degrees of freedom ($\phi_1$ and $\phi_2$) and also because the close proximity of one of the oxygen atoms of NO$_2$ to the nearest hydrogen atom on the neighbouring pyridil results in an enormous energy barrier for planar or nearly planar conformations. The effect of this large energy barrier is demonstrated in Fig.~\ref{fig1}b where we show that the total energies of the eight conformations closest to planarity are 1.5-15.0 eV higher than the one of the most stable conformation ($\phi_1$=40$^{\circ}$, $\phi_2$=50$^{\circ}$). This result is not surprising when the values of d$_{O-H}$ in the inset of Fig.~\ref{fig1}b are inspected, where for these geometries the O and H atoms in question are found to be less than 1.2 \AA \ apart in distance. Out of the remaining 92 conformations we studied in this article, however, an energy barrier of less than 0.5 eV was calculated for 85 geometries, which puts them well within the range where chemical fixation techniques as demonstrated in Refs.~\cite{hybertsen,mayor} can be employed for $\phi_1$ or where the conformation might be influenced by an external electric field~\cite{macucci}, a method that would be applicable in principle for both, $\phi_1$ and $\phi_2$. Without any form of fixation or external influence, both angles would vary over a wide range even for rather low temperatures in actual experiments due to the rather shallow energetic minima found in Fig.~\ref{fig1}b. We stress that the trends we derive and discuss for the conformation dependence of charge transfer and level alignment in our article are focused on the 92 structures that are experimentally achievable and results for the geometries close to planarity are only provided for the sake of completeness.

\section{Results and Discussion}

\subsection{Conformation dependence of the transmission function}\label{sec:trans}

In Fig.~\ref{fig2}a we show the transmission function for two selected molecular conformations in order to illustrate that a variation of $\phi_1$ can shift its peak structure close to the Fermi energy E$_F$ significantly thereby having a big impact on the conductance of the junction. We already discussed the consequences of the variation of $\phi_1$ for the observability of quantum interference effects for this system in an earlier work~\cite{qie4}, while the focus in the current article is on exploring the reasons behind the peak shifts which are a reflection of the torsion induced shifts of the MO onsite energies for the LUMO and LUMO+1 as displayed in Fig.~\ref{fig2}b. For nitro-bpph there are in principle several explanations thinkable for the torsion angle dependence of MO onsite energies: i) A change in conformation is a substantial variation of the molecular structure, and therefore has an effect on the way MOs are formed as linear combination of atomic orbitals (AOs), which in turn might alter both the energetic distances between different MOs and their spatial distribution, where the latter can have an impact on the coupling to the surface~\cite{rect5,qie4,pauly}. ii) Since the molecule is asymmetric, it exhibits a dipole moment which again might change with $\phi_1$ or $\phi_2$ and can play a role in the vacuum level alignment between molecule and leads. iii) As found for bipyridine between Au surfaces~\cite{fermimy1,fermimy2}, zero-bias charge transfer is expected to be a significant factor for level alignment also for nitro-bpph and studying its variation with the molecular conformation is therefore an important issue. It is not straightforward to separate the above named effects in a rigorous way and in the following sections we will attempt such a separation and argue that charge transfer is the dominant factor for the variation of the MO energies with $\phi_1$ and $\phi_2$ in the Au/nitro-bpph/Au system where we base this conclusion on trends in quantities we can derive directly from our DFT calculations.

\subsection{Zero-bias charge transfer and electronegativity}\label{sec:charge}
 
The starting point for our analysis has to be the question how many fractions of electrons are exchanged between the molecule and the gold surfaces at zero bias for establishing an equlibrium in terms of chemical potentials. In Ref.~\cite{fermimy1} we addressed general aspects of this question. Fractional charges are only physically meaningful in the sense of describing the charge accumulating in the immediate vicinity of a molecule adsorbed between two metal surfaces, and do not necessarily describe the filling or emptying of MOs, because these MOs are hybridizing with the metal surface states and the system cannot be partitioned into its components in an unambiguous way. The projection scheme we use in this article to derive energies for the LUMO and LUMO+1 is a way of performing such a partitioning but aimed at a special purpose. It is based on decoupling AOs on the molecule from AOs on the metal leads in the NEGF-DFT formalism, which is useful for identifing the molecular levels and their energies corresponding to peaks in the transmission function but a priori provides no information about the local charging of the molecule. 

Such information (as we established in Ref.~\cite{fermimy1}) can only be reliably retrieved in two ways: i) The lowest molecular orbital (we call it MO1 in the following) lies more than 10 eV below  all gold states and therefore is a pure molecular state, meaning that there is no hybridisation with the leads. The shift in its energy due to the interaction with the surface can therefore be only due to charging of the molecule, where the level goes up in energy when fractions of electrons are added and down if they are subtracted. A comparison of the MO1 positions in the junction and for the isolated explicitly charged molecule then allows to calculate the local charges in an indirect fashion. ii) For rather weak covalent bonds (such as the ones formed with gold by pyridil anchors as was explicitly demonstrated for Au/bipyridine/Au in Ref.~\cite{fermimy1}), the integral of the charge density differences over the spatial region of the molecule $\Delta N_{int}^{mol}$, where its density and the one of the Au slab are subtracted from the density of the combined system, also describes the charging very accurately because the border between the two systems can be drawn rather easily where the electron density has a minimum. 

We stress that charging as we define it here includes the effects of Pauli repulsion at the interface which are sometimes described as interface dipoles. Since the latter mechanism pushes fractions of electrons away from the molecule, it results in effective electron removal and pushes down the energy of all MOs in the same way as emptying the highest occupied MO (HOMO) would. This was shown in detail for Au/bipyridine/Au in Ref.~\cite{fermimy1} where Pauli repulsion was identified as dominant mechanism for the level alignment. It was also shown explicitly that even the shape of the charge density differences is the same in emptying the HOMO for the isolated bipyridine and Pauli repulsion for the Au/bipyridine/Au interface. For Au/nitro-bpph/Au the same correspondence can be found, because also for this molecule the HOMO is localized predominantly on the pyridil anchors.

  \begin{figure}
\includegraphics[width=1.0\linewidth,angle=0]{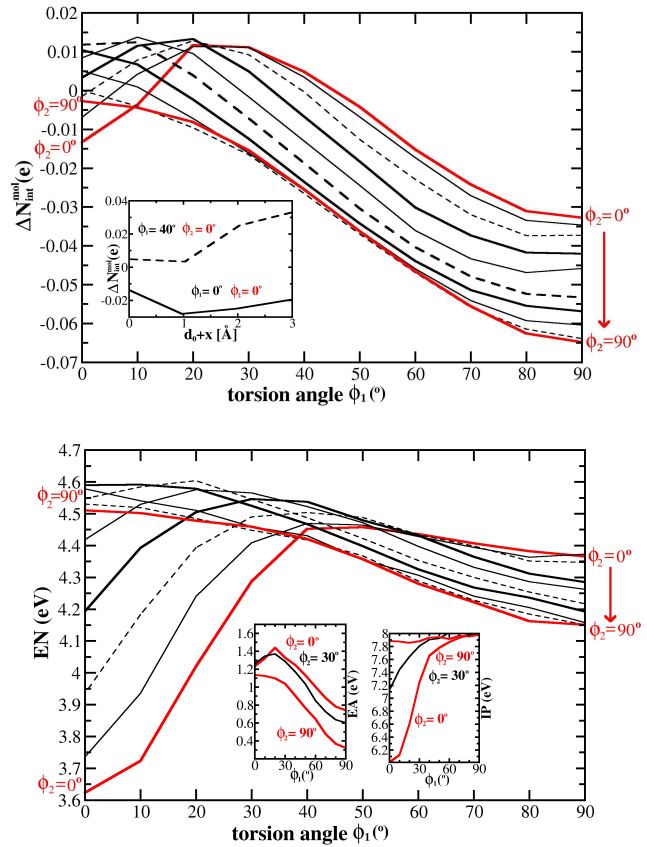}
  \caption[cap.Pt6]{\label{fig3}(Color online) (Top) Variation of the transfer of electrons from the Au surfaces to the molecule $\Delta N_{int}^{mol}$ derived from integration over charge density distribution differences between the whole junction and the molecule and surface in isolation~\cite{fermimy1} at bonding distance d$_0$ (2.12 \AA \ ) in dependence on the angles $\phi_1$ and $\phi_2$, where the x-axis denotes $\phi_1$ and different values for $\phi_2$ are distinguished by varying line types. The inset shows the distance dependence of $\Delta N_{int}^{mol}$ for the two angle combinations also highlighted in the transmission functions of Fig.~\ref{fig2}a. (Bottom) Variation of the electronegativity EN of the molecule in dependence on $\phi_1$ and $\phi_2$ with its two components, EA and IP, displayed in the insets.  }
  \end{figure}

In the top part of Fig.~\ref{fig3} we plot $\Delta N_{int}^{mol}$ in its full dependence on $\phi_1$ and $\phi_2$ for the optimal bonding distance d$_0$=2.12 \AA \ . In spite of containing NO$_2$ as a strong electron acceptor, the molecule loses fractions of electrons to the surface for a wide range of angles. The latter net result is a consequence of the interfacial Pauli repulsion discussed above. This can be illustrated by comparing the distance dependence of $\Delta N_{int}^{mol}$ for ($\phi_1$=40$^{\circ}$/$\phi_2$=0$^{\circ}$) (dashed line in the inset of the top part of Fig.~\ref{fig3}) with the study on Au/bipyridine/Au in Ref.~\cite{fermimy1}. In both cases the fractional number of electrons on the molecule goes down when its distance to the electrodes is reduced from large values to d$_0$, because the repulsion effect decays continuously with an increase of the Au-N distance. A comparison of $\Delta N_{int}^{mol}$ for d$_0$ (as shown in Fig.~\ref{fig3}) and d$_0$+3 \AA \ (not shown here) reveals that the latter effect is rather conformation independent at least for values of $\phi_1$ and $\phi_2$ higher than 30$^{\circ}$. At d$_0$+3 \AA \ the lowest value for $\Delta N_{int}^{mol}$ (for both angles higher than 10$^{\circ}$) is at ($\phi_1$=90$^{\circ}$/$\phi_2$=90$^{\circ}$), where it becomes zero. We also want to point out that the net charges on the nitro-bpph molecules are by a factor of $\sim$ 4-5 smaller than those for the unsubstituted bpph (not shown here) or bipyridine as analyzed in Ref.~\cite{fermimy1} which shows that there is a balance between the electron accepting action of NO$_2$ and interfacial Pauli repulsion explaining the relatively small amount of charging in the Au/nitro-bpph/Au system.

For a further interpretation of the trends in $\Delta N_{int}^{mol}$, it is useful to look at the torsion dependence of a related property for nitro-bpph, namely its electronegativity EN. According to Mulliken~\cite{mulliken} it can be defined as EN=(IP+EA)/2, where the ionisation potential (IP) and EA can be calculated from DFT total energies of the charged and neutral molecule as IP=E(N-1)-E(N) and EA=E(N)-E(N+1), respectively~\cite{parr}. The bottom part of Fig.~\ref{fig3} reveals that at least qualitatively the conformation dependent trends in EN match those in $\Delta N_{int}^{mol}$. From the insets it can be seen that the abrupt downward trends for low angles are mostly due to IP, while EA exhibits a continuous increase when going from high to low angles over the whole conformational range. 

The latter finding would also be expected from chemical intuition, since for high values of $\phi_2$ the $\pi$ electrons of the nitro-group and the benzene ring it is attached to are decoupled, whereas when $\phi_1$ is large the $\pi$ cloud of the inner benzene component is increasingly disconnected from those of the pyridyl anchors. Both of these disruptions of the $\pi$ connections between molecular components are bound to reduce the effect of NO$_2$ as an electron acceptor, and therefore it is in accordance with expectations that a decrease in both, $\phi_1$ and $\phi_2$, increases EA and EN, where the molecular charge $\Delta N_{int}^{mol}$ is found to follow the trend of the electron affinity as the top part of Fig.~\ref{fig3} documents at least for all angles above 20$^{\circ}$. The downward shifts of $\Delta N_{int}^{mol}$ for lower angles (also reflected in the data derived for IP) find their origin in strong steric repulsion effects for the planar molecule, where we explained the reasons in Sec.~\ref{sec:stab} and will illustrate the consequences as distortion of the electron density in Sec.~\ref{sec:dipole}. This steric repulsion is so destabilizing to nearly planar conformations  (see Fig.~\ref{fig1}b) that it puts them out of reach for actual experiments and we therefore focus in our main analysis and conclusions on angles above 20$^{\circ}$-30$^{\circ}$. 

We summarize the discussion of Fig.~\ref{fig3} by stating that a variation of the angles $\phi_1$ and $\phi_2$ can alter the fractional charge on the molecule by $\sim$ 0.1 electrons. Due to the strong Pauli repulsion identified for the pyridyl/Au bond in earlier work~\cite{fermimy1}, $\Delta N_{int}^{mol}$ is negative for a wide range of the investigated conformations, which means that nitro-bpph becomes positively charged in this junction. The conformation dependence of $\Delta N_{int}^{mol}$ is due to a decoupling of $\pi$ electrons on molecular components for increased torsion angles, which is also reflected in values for EN and EA calculated for the molecule in isolation. We stress that no dependence of $\Delta N_{int}^{mol}$ on $\phi_1$ exists for bpph without a nitro-group (not shown here) which is further support for our arguments.

\subsection{Molecular level positions}\label{sec:levels}
  \begin{figure*}
\includegraphics[width=0.7\linewidth,angle=270]{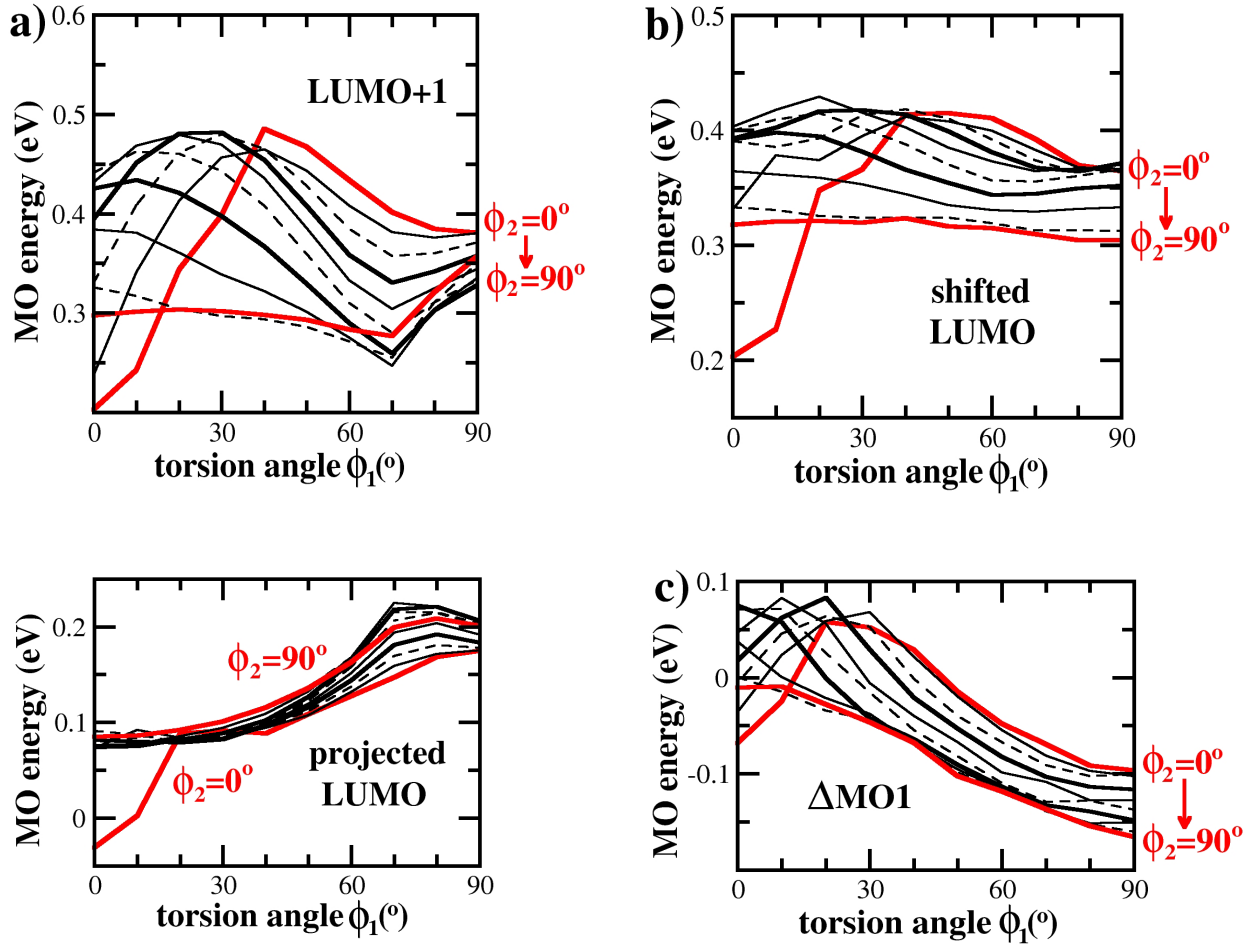}
\vspace{0.5 cm}
\caption[cap.Pt6]{\label{fig4}(Color online) a) Torsion dependence of the eigenenergies for the LUMO and LUMO+1 as derived from a projection scheme for molecular levels within the NEGF formalism~\cite{kristian} from the electron density of the junction by subdiagonalization of the Hamiltonian built from the basis functions localized on the molecule~\cite{mikkel}. b) Variation of LUMO energies with $\phi_1$ and $\phi_2$ as obtained by adding the energetic difference between the LUMO and the lowest molecular level MO1 for the isolated molecule to the easily identifiable MO1 energy within the composite junction. c) Angle dependent variation of the shift of MO1 for the isolated molecules due to fractional charging corresponding to the values in Fig.~\ref{fig3}. All energy values in this figure are given with respect to the Fermi energy of the leads. }
  \end{figure*}

After having addressed the conformation dependence of zero bias charge transfer, the next step of our analysis is an evaluation of its impact on the alignment of molecular levels with respect to E$_F$. One way of defining MOs within the junction in spite of the coupling between molecule and electrodes is provided by the NEGF formalism~\cite{kristian}, where the Hamiltonian for the electron transport is formulated in terms of the localized Siesta basis functions~\cite{siesta} and a subdiagonalization of the molecular AOs only can be performed~\cite{mikkel}. This method gives MO eigenenergies incorporating all electrostatic interactions between the molecule and the leads including charge transfer but not direct hybridisation between AOs on separate components of the junction. We show the results of this projection scheme for the LUMO and LUMO+1 in Fig.~\ref{fig4}a in dependence on both $\phi_1$ and $\phi_2$. While the level positions for the LUMO+1 roughly follow $\Delta N_{int}^{mol}$ and EN (as shown in Fig.~\ref{fig3}) in their dependence on molecular conformation, the projected values for the LUMO differ considerably in their trends.

In order to investigate this discrepancy further we employ another way for predicting the LUMO energy in Fig.~\ref{fig4}b. Here we make use of the fact that the energetic position of the lowest lying molecular orbital MO1 is more than 10 eV below all of the Au valence states and is therefore independent of the hybridisation between the molecule and the leads~\cite{fermimy1}. This makes it possible to read the MO1 energy directly from the standard output of the DFT calculations for the whole system and add the energetic difference between MO1 and the LUMO as calculated for the molecule in isolation, where both has been obtained independently for all values of $\phi_1$ and $\phi_2$. The resulting predictions in Fig.~\ref{fig4}b do not take into account any form of direct interaction between the LUMO and Au states but they do include all electrostatic effects such as charge tranfer and changes in the vacuum level alignment due to modified dipole moments (see Sec.~\ref{sec:dipole} for a discussion of the dependence of the molecular dipole moment on the torsion angles) as well as variations of the energy differences between MOs in the isolated molecule due to the distortion. Since these predictions show a picture that is in agreement in terms of angle dependent trends with the LUMO+1 but not the LUMO in Fig.~\ref{fig4}a, we can conclude that the LUMO positions from the projection scheme must be governed by the interaction of this particular MO with surface states. This finding can be explained by the observation that while the molecule can lose electrons due to Pauli repulsion at the pyridyl/Au interface, the only way it has of obtaining fractions of electrons from the Au surfaces is to have its LUMO partially occupied (which would automatically bring it closer to E$_F$) and that this effect should increase with a decrease of both torsion angles (see the discussion in Sec.~\ref{sec:charge}) as indeed can be seen in the lower part of Fig.~\ref{fig4}a.

Finally, we want to isolate the effect that charge transfer has on molecular level positioning from other conformation dependent quantities such as the molecular dipole moment and the MO structure of the molecule. For this purpose we perform two separate sets of DFT calculations on the isolated molecule, where for one of them we define a charge neutral setup and for the other we add the fractional charges~\cite{casida} from Fig.~\ref{fig3} to the total number of electrons in the unit cell. Assuming that these charges are so small that their impact on the position of the vacuum potential in the calculation is negligible (which appears to be reasonable because the dipole moment remains unaltered by the charging), we argue that by reading the energy MO1 for both sets and forming the difference $\Delta$MO1 between them in dependence on $\phi_1$ and $\phi_2$ as plotted in Fig.~\ref{fig4}c, we get a measure of the level variation that is caused by charge transfer only. This variation is very comparable both in its scope of energies (0.2-0.3 eV) and its trends with the torsion angles to the conformation dependencies of the LUMO in the prediction of Fig.~\ref{fig4}b and the LUMO+1 in the projection scheme of Fig.~\ref{fig4}a. From this finding we conclude that charge transfer is the dominant source for the conformation dependence of the level alignment in the Au/nitro-bpph/Au system, because neither the molecular dipole moment nor a changed MO structure can have an impact on the observed variations of $\Delta$MO1 with $\phi_1$ and $\phi_2$ in Fig.~\ref{fig4}c. The charge induced behaviour of $\Delta$MO1, however, mimics the trends in the angle dependent shift of the most relevant peaks in the transmission function in Fig.~\ref{fig2} which is a central result of our article.  

\subsection{Scissor operator correction for the shortcomings of DFT in describing MO energies}\label{sec:scissor}

\begin{figure} 
\includegraphics[width=1.0\linewidth,angle=0]{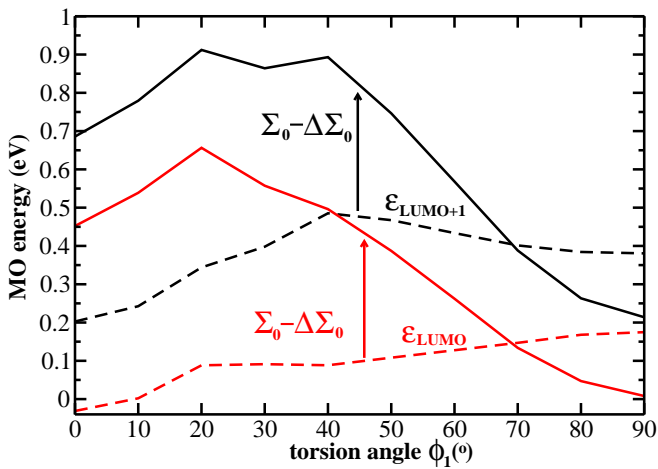}
\caption[cap.Pt6]{\label{fig5}(Color online) Shift of the levels $\epsilon_{LUMO}$ and $\epsilon_{LUMO+1}$ in Fig.~\ref{fig2} due to the {\it ad hoc} correction with $\Sigma_0 - \Delta \Sigma_0$ introduced in Ref.~\cite{shift1}, where uncorrected values are plotted with dashed and corrected values with solid lines.}
\end{figure}

It is well known that the single electron energy gaps in DFT are too small when compared to a many body description for insulators and semiconductors (including finite molecules) due to self-interaction effects and the approximate description of electronic correlations~\cite{kristian1}. In addition the gap reduction due to screening effects in metallic leads (which are often described by classical image charge models~\cite{shift1,ratner}) is also not contained in transmission functions calculated from NEGF-DFT at zero bias~\cite{kristian2}. Recently, it has been shown that both errors cancel out to a high extent at least for the occupied MOs of benzene molecules adsorbed on metal surfaces~\cite{kristian3}. Our study addresses the relative conformation dependent changes of MO energies and their relation to zero bias charge transfer~\cite{fermimy1} within the approximations of a single particle NEGF-DFT approach, where so far we have assumed that the described shortcomings of our method are not a major concern. This assumption was based mostly on the argument that the calculated transmission functions can be compared with experiments due to the error cancellation cited above. We stress that the screening effects in the leads due to the molecular dipole or interface dipole moments are properly described within DFT and that it is only the polarisation of the metal electrons due to electronic excitations of molecular states which are addressed in Ref.~\cite{kristian2}.

It is outlined, however, in Ref.~\cite{kristian3} that, although there will always be some amount of cancellation between self-interaction errors and missing polarization effects in a semi-local DFT description, the relative size of the two contributions might depend significantly on a variety of factors such as the structure of the molecule, its orientation with respect to the surface, the molecule-surface distance and the type of substrate. Since some of these aspects differ strongly when we compare the systems in our work with those of Ref.~\cite{kristian3}, we want to investigate both sources of error explicitly in the following. For this purpose and applying the procedure for an {\it ad hoc} correction in Ref.~\cite{shift1}, we introduce a scissor operator $\Sigma_0 - \Delta \Sigma_0$, where $\Sigma_0 = -(\epsilon_{LUMO} + EA)$ emends the energy gap error for the isolated molecule and $\Delta \Sigma_0$ accounts for dynamical polarisation effects due to the presence of the metal surface as calculated from an image charge model (see the appendix of Ref.~\cite{shift1} for the details of this approach). In Fig.~\ref{fig5} we plot the uncorrected and corrected $\epsilon_{LUMO}$ and $\epsilon_{LUMO+1}$ in their dependence on $\phi_1$ for $\phi_2$=0$^{\circ}$. The following observations can be made: i) The two contributions $\Sigma_0$ and $\Delta \Sigma_0$ cancel out exactly only for one geometry ($\phi_1$=70$^{\circ}$). In general $\Sigma_0$ varies quite significantly with $\phi_1$ (see also the inset of Fig.~\ref{fig3}b for the $\phi_1$ dependence of EA), while $\Delta \Sigma_0 = 0.75 \pm 0.03$ for all structures; ii) although the scissor operator correction alters the energies of both levels quite strongly in quantitative terms, the angle dependent trends discussed in the sections above are not only still observable, but their observability is even enhanced by the correction, where also the corrected levels rise in energy with a decrease of $\phi_1$ for all angles higher than 20-30$^{\circ}$.

\subsection{Molecular dipole moments}\label{sec:dipole}

  \begin{figure}
  \includegraphics[width=1.0\linewidth,angle=0]{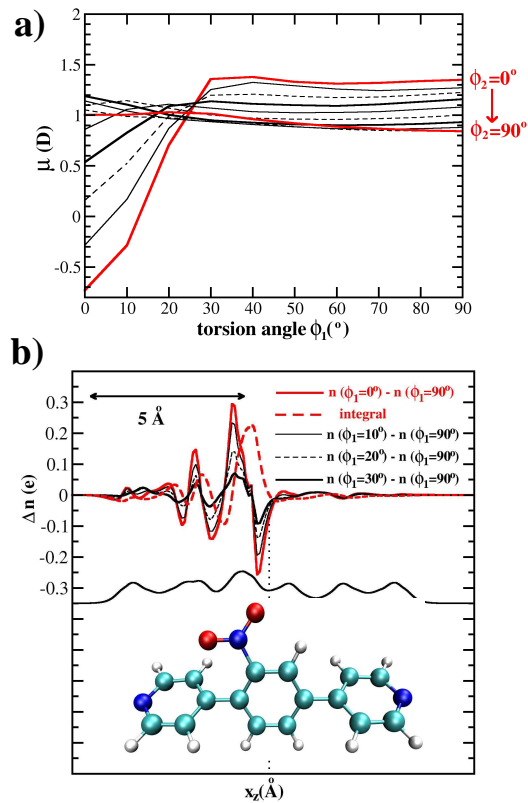}
  \caption[cap.Pt6]{\label{fig6}(Color online) a) Molecular dipole moments $\mu$ for nitro-bpph in dependence on the torsion angles $\phi_1$ and $\phi_2$. b) With $\phi_2$ fixed to 0$^{\circ}$, the effect of steric repulsion on the charge density is investigated for nearly planar molecules, where the density for $\phi_1$=90$^{\circ}$ is taken as a reference and the sum of the charge density is formed over the plane perpendicular to the axis defined by the two pyridil-N-atoms. The inset shows the structure of the molecule, where the evolution of the total electron density along this axis is also indicated in arbitrary units.  }
  \end{figure}

For molecules without a dipole moment in equal distance to two identical surfaces, the alignment of molecular levels is completely defined by a balance of Pauli repulsion at the interface, which is the cause for the formation of interface dipoles in such a setup and the dominant source for electron subtraction from the molecule, and a partial filling of the LUMO, which is the only possibility for the molecule to gain electrons. The interplay of these two driving forces for level alignment has been investigated for bipyridine and biphenyl-dithiolate between Au leads~\cite{fermimy1,fermimy2}. If the junction is asymmetric, the picture becomes more complicated. While it has been shown that for monolayers of molecules adsorbed on just one surface the level alignment is dominated by interface dipoles, while molecular dipoles only affect the work function~\cite{heimel1,heimel2}, no such rigorous separation of effects can be motivated for molecules with an inherent dipole in the "sandwhich" setup of the junctions we study in this article. We face the additional complication that vacuum level alignment is also not straightforward in such a junction, because two different vacuum levels exist to the left and right of the molecule due to the dipole. Throughout this article we avoid the latter problem, since we only investigate the dependence of level positions on the torsion angles $\phi_1$ and $\phi_2$, thereby deriving indirect arguments for the dominance of charge transfer effects, but do not decompose the level alignment explicitly for a given conformation in terms of the relative strength of contributing factors.

In Fig.~\ref{fig6}a we show the molecular dipole moment $\mu$ as calculated from the electron density and ion distributions~\cite{siesta} for all molecular conformations. Strikingly, there does not seem to be any dependence of $\mu$ on $\phi_1$ above 30$^{\circ}$, while only for $\phi_2$ there are trends which can compare to those for charge transfer and level alignment in the previous sections. This result seems plausible, because the angle $\phi_1$ determines how much the inner benzene ring can attract electrons from the two pyridil groups, where there is no reason to believe that this charge transfer phenomen should exhibit a distinct asymmetry which would then induce a change of the dipole moment. The second angle $\phi_2$, on the other hand, determines how many fractions of electrons the nitro-group receives from the whole chain of aromatic rings. Since the partial charges on NO$_2$ are what causes the molecular dipole moment, the relation of $\mu$ to $\phi_2$ becomes evident. This finding is additional proof that at least the dependence of level alignment on $\phi_1$ is caused by the change in charge transfer between the molecule and surfaces, simply because $\mu$ does not change with $\phi_1$.

Finally, we take a closer look at what causes the dramatic change of $\mu$ for $\phi_1$ below 30$^{\circ}$ by forming charge density differences for these molecular conformations with reference to a molecule with perpendicular aromatic rings (see Fig.~\ref{fig6}b). It can be seen that the more the conformation approaches planarity, the more electrons are pushed away from the region where the H and O atoms highlighted in Fig.~\ref{fig1}a meet and partly pushed away from the molecule altogether as seen in Fig.~\ref{fig3}a. Interestingly, another net result of this repulsion effect, is a shift of electrons towards the right side of the molecule (where there is no nitro-group), thereby reducing the molecular dipole moment and ultimatively (for $\phi_1$=0$^{\circ}$ and 10$^{\circ}$) reversing its direction.

\section{Conclusions}

In summary, we demonstrated that the shifts of molecular levels and peaks in the transmission functions in the Au/nitro-bpph/Au system induced by a variation of molecular conformation can be explained in terms of zero bias charge transfer which in turn is governed by the decoupling of $\pi$ electrons between molecular components with an increase in torsion angles. The electron accepting nature of NO$_2$ groups in this junction is countered by Pauli repulsion at the interface between the pyridil anchors and Au leads, where the latter effect does not depend on conformation. 

These findings are relevant for the evaluation of the practicability of theoretical device proposals based on nitro benzene where it is usually taken for granted that adding NO$_2$ to an aromatic ring always induces negative partial charges on the molecule and such dependencies on molecular conformation and anchor groups are in general not discussed or even considered. 

While the torsion angle dependence of the electronic coupling in molecules without acceptor groups and rather strongly coupled to gold leads has been studied previously~\cite{rect5,pauly,lambert}, the effect of molecular conformation on level alignment in weaker coupled systems and containing acceptor groups has so far not been addressed. The two cases differ entirely in the respective mechanisms for the conformation dependence. 

Since very recently chemical means have been found to fix torsion angles in actual experiments~\cite{hybertsen,mayor}, our results might open up an avenue for a further exploration of effects related to molecular conformation in single molecule junctions and even new device schemes for molecular switches might result from such experiments. 

\begin{acknowledgements}

The author is currently supported by the Austrian Science Fund FWF, project Nr. P20267. He would like to thank Victor Geskin for his invaluable advice regarding the calculation of the screening correction to molecular orbital eigenenergies from an image charge model. Helpful discussions with Jan Zabloudil are gratefully acknowledged.

\end{acknowledgements}


\bibliographystyle{apsrev}

\begin{thebibliography}{23}
\expandafter\ifx\csname natexlab\endcsname\relax\def\natexlab#1{#1}\fi
\expandafter\ifx\csname bibnamefont\endcsname\relax
  \def\bibnamefont#1{#1}\fi
\expandafter\ifx\csname bibfnamefont\endcsname\relax
  \def\bibfnamefont#1{#1}\fi
\expandafter\ifx\csname citenamefont\endcsname\relax
  \def\citenamefont#1{#1}\fi
\expandafter\ifx\csname url\endcsname\relax
  \def\url#1{\texttt{#1}}\fi
\expandafter\ifx\csname urlprefix\endcsname\relax\def\urlprefix{URL }\fi
\providecommand{\bibinfo}[2]{#2}
\providecommand{\eprint}[2][]{\url{#2}}

\bibitem{molelect}G. Cuniberti, G. Fagas, K. Richter, Introducing Molecular Electronics, {\it Lect. Notes Phys.} {\bf 680} (Springer, New York, 2005).
\bibitem{h2pt1}R. H. M. Smit, Y. Noat, C. Untiedt, N. D. Lang, M. C. van Hemert and J. M. van Ruitenbeek, {\it Nature} {\bf 419}, 906 (2002).
\bibitem{h2pt2}D. Djukic, K. S. Thygesen, C. Untiedt, R. H. M. Smit, K. W. Jacobsen and J. M. van Ruitenbeek, {\it Phys. Rev. B} {\bf 71}, 161402(R) (2005).
\bibitem{h2pt3}D. Djukic and J. M. van Ruitenbeek, {\it Nano Lett.} {\bf 6}, 789 (2006). 
\bibitem{h2pt4}M. Kiguchi, R. Stadler, I. S. Kristensen, D. Djukic and J. M. van Ruitenbeek, {\it Phys. Rev. Lett.} {\bf 98}, 146802 (2007).
\bibitem{bipy1}B. Xu and N. J. Tao, {\it Science} {\bf 301}, 1221 (2003).
\bibitem{bipy2}B. Q. Xu, X. Y. Xiao and N. J. Tao, {\it J. Am. Chem. Soc.} {\bf 125}, 16164 (2003).
\bibitem{bipy3}R. Stadler, K. S. Thygesen and K. W. Jacobsen, {\it Phys. Rev. B} {\bf 72}, 241401(R) (2005).
\bibitem{bipy4}S. Hou, J. Zhang, R. Li, J. Ning, R. Han, Z. Shen, X. Zhao, Z. Xue, and Q. Wu, {\it Nanotechnology} {\bf 16}, 239 (2005).
\bibitem{bipy5}A. J. Perez-Jimenez, {\it J. Phys. Chem. B} {\bf 109}, 10052 (2005).
\bibitem{alkdithio1}W. Y. Wang, T. Lee and M. A. Reed, {\it Phys. Rev. B} {\bf 68}, 035416 (2003).
\bibitem{alkdithio2}C. C. Kaun, and H. Guo, {\it Nano Lett.} {\bf 3}, 1521 (2003).
\bibitem{alkdithio3}W. Y. Wang, T. Lee, I. Kretzschmar, and M. A. Reed, {\it Nano Lett.} {\bf 4}, 643 (2004).
\bibitem{alkdithio4}T. Lee, W.Y. Wang, J. F. Klemic, J. J. Zhang, J. Su and M. A. Reed, {\it J. Phys. Chem. B} {\bf 108}, 8742 (2004).
\bibitem{alkdithio5}V. B. Engelkes, J. M. Beebe, and C. D. Frisbie, {\it J. Am. Chem. Soc.} {\bf 126}, 14287 (2004).
\bibitem{alkdithio6}H. B. Akkerman, P. W. M. Blom, D. M. de Leeuw and B. de Boer, {\it Nature} {\bf 441}, 69 (2006).
\bibitem{device1}C. Joachim, J. K. Gimzewski and A. Aviram, {\it Nature} {\bf 408}, 541 (2000).
\bibitem{device2}J. C. Ellenbogen and J. C.  Love, {\it Proc. IEEE} {\bf 88}, 386 (2000).
\bibitem{device3}R. M. Metzger, {\it J. Mater. Chem.} {\bf 18}, 4364 (2008).
\bibitem{device4}J. Ferrer and V. M. Garcia-Suarez, {\it J. Mater. Chem.} {\bf 19}, 1696 (2009).
\bibitem{rect1}K. Stokbro, J. Taylor and M. Brandbyge, {\it J. Am. Chem. Soc.} {\bf 125}, 3674 (2003).
\bibitem{rect2}A. Staykov, D. Nozaki and K. Yoshizawa, {\it J. Phys. Chem. C} {\bf 111}, 11699 (2007).
\bibitem{rect3}R. Stadler, V. Geskin and J. Cornil, {\it Adv. Funct. Mater.} {\it 18}, 1119 (2008).
\bibitem{rect4}R. Stadler, V. Geskin and J. Cornil, {\it J. Phys.: Condens. Matter} {\it 20}, 374105 (2008).
\bibitem{rect5}H. Kondo, J. Nara, H. Kino, and T. Ohno, {\it J. Chem. Phys.} {\bf 128}, 064701 (2008).
\bibitem{qie1}R. Stadler, K. S. Thygesen and K. W. Jacobsen, {\it Nanotechnology}, {\bf 16}, S155 (2005).
\bibitem{qie2}D. Q. Andrews, G. C. Solomon, R. P. Van Duyne and M. A.  Ratner, {\it J. Am. Chem. Soc.} {\bf 130}, 17309-17319 (2008).
\bibitem{qie3}G. C. Solomon, D. Q. Andrews, R. P. Van Duyne and M. A. Ratner, {\it ChemPhysChem} {\bf 10}, 257 (2009).
\bibitem{qie4}R. Stadler, {\it Phys. Rev. B} {\bf 80}, 125401 (2009).
\bibitem{scandolo}R. Rousseau, V. De Renzi, R. Mazazzarello, D. Marchetto, R. Biagi, S. Scandolo and U. del Pennino, {\it J.Phys. Chem. B} {\bf 110}, 10862-10872 (2006).
\bibitem{heimel1}G. Heimel, L. Romaner, J. L. Bredas and E. Zojer, {\it Phys. Rev. Lett.} {\bf 96}, 196806 (2006).
\bibitem{heimel2}G. Heimel, L. Romaner, E. Zojer and J. L. Bredas, {\it Nano Lett.} {\bf 7}, 932 (2007).
\bibitem{fermimy1}R. Stadler and K. W. Jacobsen, {\it Phys. Rev. B} {\bf 74}, 161405(R) (2006).
\bibitem{fermimy2}R. Stadler, {\it J. Phys.: Conf. Ser.} {\bf 61}, 1097 (2007).
\bibitem{flores}H. Vazquez, Y. J. Dappe, J. Ortega and F. Flores, {\it Appl. Surf. Sci.}, {\bf 254}, 378 (2007).
\bibitem{brocks}P. C. Rusu, G. Giovannetti and G. Brocks, {\it J. Phys. Chem. C} {\bf 111}, 14448 (2007).
\bibitem{heimel3}G. Heimel, E. Zojer, L. Romaner, J. L. Bredas and F. Stellaci, {\it Nano Lett.} {\bf 9}, 2559 (2009). 
\bibitem{shift1}D. J. Mowbray, G. Jones and K. S. Thygesen, {\it J. Chem. Phys.} {\bf 128}, 111103 (2008).
\bibitem{shift2}M. Smeu, R. A. Wolkow and G. A. DiLabio, {\it J. Chem. Phys.} {\bf 129}, 034707 (2008).
\bibitem{hybertsen1}L. Venkataraman, Y. S. Park, A. C. Whalley, C. Nuckolls, M. S. Hybertsen and M. L. Steigerwald, {\it Nano Lett.} {\bf 2}, 502 (2007).
\bibitem{ratner}C. Risko, C. D. Zangmeister, Y. Yao, T. J. Marks, J. M. Tour, M. A. Ratner and R. D.  van Zee, {\it J. Phys. Chem. C} {\bf 112}, 13215 (2008).
\bibitem{siesta}J. M. Soler, E. Artacho, J. D. Gale, A. Garcia, J. Junquera, P. Ordejon and D. Sanchez-Portal, {\it J. Phys.: Cond. Matt.} {\bf 14}, 2745 (2002).
\bibitem{hybertsen}L. Venkataraman, J. E. Klare, C. Nuckolls, M. S. Hybertsen, M. L. Steigerwald, {\it Nature} {\bf 442}, 904 (2006).
\bibitem{mayor}A. Shaporenko, M. Elbing, A. Blaszczyk, C. von H\"{a}nisch, M. Mayor and M. Zharnikov, {\it J. Phys. Chem. B} {\bf 110}, 4307 (2006).
\bibitem{pauly}F. Pauly, J. K. Viljas, J. C. Cuevas and G.  Sch\"{o}n, {\it Phys. Rev. B} {\bf 77}, 155312 (2008).
\bibitem{lambert}C. M. Finch, S. Sirichantaropass, S. W. Bailey, I. M. Grace, V. M. Garcia-Suarez and C. J. Lambert, {\it J. Phys.: Condens. Matter} {\bf 20}, 022203 (2008).
\bibitem{kristian}K. S. Thygesen, and K. W. Jacobsen, {\it Chem. Phys.} {\bf 319}, 111 (2005).
\bibitem{mikkel}M. Strange, I. S. Kristensen, K. S. Thygesen and K. W. Jacobsen, {\it J. Chem. Phys.} {\bf 128}, 114714 (2008).
\bibitem{mulliken}R. S. Mulliken, {\it J. Chem. Phys.} {\bf 2}, 782-793 (1934).
\bibitem{parr}R. G. Parr, R. A. Donelly, M. Levy and W. E. Palke, {\it J. Chem. Phys.} {\bf 68}, 3801 (1977).
\bibitem{casida}M. E. Casida, {\it Phys. Rev. B} {\bf 59}, 4694 (1999).
\bibitem{martins}N. Troullier and J. L. Martins, {\it Solid State Commun.} {\bf 74}, 613 (1990).
\bibitem{perdew2}J. P. Perdew, K. Burke and M. Ernzerhof, {\it Phys. Rev. Lett.} {\bf 77}, 3865 (1996).
\bibitem{keldysh}Y. Meir and N. S. Wingreen, {\it Phys. Rev. Lett.} {\bf 68}, 2512 (1992).
\bibitem{ordejon}S. Garcia-Gil, A. Garcia, N. Lorente and Pablo Ordejon, {\it Phys. Rev. B} {\bf 79}, 075441 (2009).
\bibitem{biphenyl}L. F. Pacios and L. Gomez, {\it Chem. Phys. Lett.} {\bf 432}, 414 (2006).
\bibitem{macucci}I. Cacelli, A. Ferretti, M. Girlanda and M. Macucci, {\it Chem. Phys.} {\bf 320}, 84 (2006).
\bibitem{kristian1}K. S. Thygesen, {\it Phys. Rev. Lett.} {\bf 100}, 166804 (2008).
\bibitem{kristian2}K. S. Thygesen and A. Rubio, {\it Phys. Rev. Lett.} {\bf 102}, 046802 (2009).
\bibitem{kristian3}J. M. Garcia-Lastra, C. Rostgaard, A. Rubio, and K. S. Thygesen, {\it Phys. Rev. B} {\bf 80}, 245427 (2009).
\end{thebibliography}

\end{document}